\documentclass[11pt]{article}
\usepackage{mathrsfs}
\usepackage{amsxtra}
\usepackage{stmaryrd}
\usepackage{psfrag}
\usepackage{subfigure}
\usepackage{rotating} 
\usepackage{natbib}
\usepackage{color}

\begin{document}
\title{In-Plane fracture of laminated fiber reinforced composites with varying fracture resistance: experimental observations and numerical crack propagation simulations}
\date{}
\author{S.~Rudraraju\thanks{Department of Mechanical Engineering, University of Michigan, Ann Arbor, {\tt rudraa@umich.edu}}, A.~Salvi\thanks{Department of Aerospace Engineering, University of Michigan, Ann Arbor, {\tt amits@umich.edu}} , K.~Garikipati\thanks{Department of Mechanical Engineering, University of Michigan, Ann Arbor, {\tt krishna@umich.edu}} \& A.M.~Waas\thanks{Department of Aerospace Engineering, University of Michigan, Ann Arbor, corresponding author, {\tt dcw@umich.edu}}}
\maketitle
\abstract{A series of experimental results on the in-plane fracture of a fiber reinforced laminated composite panel is analyzed using the variational multi-scale cohesive method (VMCM). The VMCM results demonstrate the influence of specimen geometry and load distribution on the propagation of large scale bridging cracks in the fiber reinforced panel. Experimentally observed variation in fracture resistance is substantiated numerically by comparing the experimental and VMCM load-displacement responses of geometrically scaled single edge-notch three point bend (SETB) specimens. The results elucidate the size dependence of the traction-separation relationship for this class of materials even in moderately large specimens, contrary to the conventional understanding of it being a material property. The existence of a ``free bridging zone'' (different from the conventional ``full bridging zone'') is recognized, and its influence on the evolving fracture resistance is discussed. The numerical simulations and ensuing bridging zone evolution analysis demonstrates the versatility of VMCM in objectively simulating progressive crack propagation, compared against conventional numerical schemes like traditional cohesive zone modeling, which require a priori knowledge of the crack path.}

\section{Introduction}
\label{introduction}

Laminated fiber reinforced composites are finding increased use as structural materials in a variety of aerospace and other industrial applications.  An appealing property of these composites is their high specific strength (strength per unit weight). Even though a strong case can be made for using composite structural parts in several areas of a load bearing structure, the requirement to demonstrate structural integrity and damage tolerance (SIDT) necessitates a fundamental understanding of the mechanical response, damage tolerance and damage growth of a load bearing composite structure. While several investigations have addressed damage growth in the form of delamination crack growth, relatively little has been done to understand damage propagation when a crack, or damage in the form of a crack that has severed all laminae of a laminate, is present in a composite structure. The influence of such wide area damage on the load bearing ability of a homogeneous and isotropic material has received considerable attention in the past; however, a similar effort at understanding issues in a non-homogeneous and macroscopically orthotropic structure is still a problem that requires resolution. Because of the different length scales associated with the microstructure of a composite material and the resulting composite structure, a multitude of failure mechanisms can be simultaneously operative, leading to a very complex damage progression in a composite structure. A sharp, through the thickness crack can be present in these composites initially, but, as soon as local damage (this can be in the form of matrix micro-cracking) accumulates, crack blunting and distributed damage occurs across the highly stressed areas around the initial crack tip. As this initial crack starts to grow, a damaged zone of material (bridging zone) evolves in the wake of the instantaneous crack tip. Thus, unlike in monolithic materials (such as metals), there is no well defined ``crack'' that can be identified. Instead, a diffused zone of damage is seen to advance. This distributed damage results in additional resistance to advancing damage growth, largely contributed by fiber bridging and pullout in the crack wake . This enhanced fracture resistance is desirable and is a major contributor to the increased toughness of these laminated composites (\citet{Cooper1970}, \citet{Kelly1971}, \citet{Kelly1973}, \citet{Cox1991}).\\

Analytical models based on linear elastic fracture mechanics (LEFM) have been developed and
implemented within finite element codes to study a variety of fracture problems. LEFM based approaches have proven to be effective in predicting crack initiation and subsequent growth in cases where material nonlinearity is negligible, and process zones are small (\citet{Hertzberg1983}, \citet{Xie2006}, \citet{DXie2006}, \citet{Waas2004}, \citet{Waas2005}). However, in heterogeneous materials, like laminated fiber composites, the process zone size may be larger than any characteristic problem length scale, and thereby, the  basic tenets of LEFM cease to hold (\citet{Amit2006}).
Several mechanisms may contribute to this situation. Micro-cracking, fiber bridging, coalescence of voids and
other microstructural mechanisms can give rise to a process zone that is considerably larger than that permitted
for the application of LEFM models. Furthermore, the material non-linearity that is induced by these
mechanisms leads to a relief of the singular fields that would otherwise persist in a strict LEFM setting of an elastic material. A new length scale, $l^\ast$, emerges that is related to a characteristic elastic modulus $E$,
fracture toughness $\Gamma$ and cohesive strength, $\sigma_c$, defined as, ${l^\ast}^2={{E\Gamma}\over {\sigma_c^2}}$.
If $l^\ast$ is larger than any characteristic length scale in the problem, then, cohesive zone models (CZM) become an
indispensable tool for analysis (\citet{Mroz1981}, \citet{Knauss1987}, \citet{Hutchinson1992}, \citet{Borst1993}, \citet{Needleman1994}, \citet{Ortiz1996}).\\

In order to implement a CZM in its simplest form, two fracture parameters are required: a fracture toughness (or energy) and a cohesive strength. Both parameters can be experimentally determined by coupon level tests, and are subsequently used as material properties for prediction of crack growth in other structural configurations. In CZM, an existing crack starts to grow when the crack tip stresses reach the cohesive strength, and evolves according to the energy available in the deformed system to create additional cracked surface. Based on these two parameters, a cohesive traction-separation law is assumed in numerical simulations of crack growth. This micro-mechanical law relates the evolution of crack face tractions with the crack face opening displacement, and numerically manifests the resistance offered to crack advancement within the cohesive zone.  This two
parameter dependent evolution, unlike the solely fracture toughness based LEFM schemes, better represents the physics of crack growth in materials with significant process zones, and it has been widely used in many numerical implementations, like the discrete cohesive zone method (DCZM) (\citet{Waas2006}, \citet{Amit2006}). However, all implementations of CZM methodology require a priori knowledge of the ``intended'' crack path. This prevents CZM from being applied to a
wide range of problems involving arbitrary crack path evolution. Conventional finite element based implementations of CZM (\citet{Ortiz2001}) constrain the crack path evolution to element interfaces, as standard finite elements lack the ability to
represent cracks within the element domain. This restriction places a limitation on those problems for which the crack path direction is not known a priori.\\

In addition to the above drawbacks of CZM models, recent work by \citet{Sun2005}, \citet{Sun2006}, has addressed some basic issues pertaining to the reconciliation of LEFM and CZM. In particular, two issues have been considered; (1) when CZM is used, the placement of CZM elements along the intended crack path can lead to an alteration of the stiffness of the original body that is to be studied, and, (2) The traction-separation laws used for traditional CZM modeling, which start with a vanishing traction at vanishing separation, may be in conflict with the presence of an intense stress field that was present in the original body that is being modeled.\\

To order to address these basic issues and to circumvent the numerical restrictions on unrestricted and objective simulation of crack propagation in materials, a micro-mechanics based, mesh independent numerical technique for simulating crack propagation is essential. Standard finite elements fail to accomplish this task as they lack the ability to capture the discontinuous displacement modes involved in crack propagation problems. However, in recent years finite elements with discontinuities (enhanced finite elements) have gained increasing interest in modeling material failure, due to their
ability to capture the specific kinematics of a displacement discontinuity(like cracks) through additional discontinuous deformation modes. In discontinuous displacement enhanced finite elements, the crack path is present inside the elements, unlike cohesive zone methods which are restricted to crack propagation along element interfaces. The ability of the enhanced finite element to encompass a crack path, leads to objective simulation of crack propagation without mesh bias. Depending on the support of the enriching discontinuous displacement modes, the enhanced finite elements are popularly classified as element enrichment methods (\citet{Armero1996}, \citet{Garikipati1998}, \citet{Jirasek2000}, \citet{Borja2000}, \citet{Huespe2004}, \citet{Mosler2004}, \citet{Holzapfel2003}) and nodal enrichment methods (XFEM, \citet{Belytschko2001}, \citet{Belytschko2003}, \citet{Sukumar2000}, \citet{Wells2001}). Interested readers are referred to \citet{Oliver2006}, for detailed discussion and comparison of these methods. Though these enhanced methods provide a general numerical framework for simulation of crack evolution, the actual micromechanics implementation which incorporates the physics of crack formation is wide open. In this context, we present the Variation Multiscale Cohesive Method (VMCM), which is an enhanced finite element method containing elemental displacement field enrichment, naturally arising out of the variational multiscale formulation presented in \citet{Garikipati1998}, \citet{Garikipati2002}, seamlessly embedding the cohesive nature of crack path evolution. In this paper, the VMCM method advanced by the authors is briefly presented, and it is used to study through the thickness crack propagation in fiber reinforced laminated panels. A more detailed presentation of VMCM is available in other related studies by the authors (\citet{Garikipati2002}, \citet{Rudraraju2008}), where the method is referred to as the variational multiscale method (VMM).\\

The paper is organized as follows: In Section ~\ref{experiments}, the application of VMCM to study crack propagation is motivated through a discussion of experimental results in scaled SETB specimens, loaded to failure. Section ~\ref{VMCM} presents the details of VMCM implementation using the finite element method. Simulations of SETB tests with varying specimen size are provided in Section ~\ref{simulations}. Comparison of experimental results, against the predictions of VMCM are discussed in Section ~\ref{discussion}, while concluding remarks are presented in Section ~\ref{conclusion}.


\section{Motivation}
\label{experiments}
The primary motivation for the current work is the experimentally observed scaling in the in-plane fracture resistance,\footnote{Traditionally, fracture resistance is defined in terms of the energy released per unit area of crack surface formation.  But, unlike in most monolithic materials, there is no well-defined ``crack'' that can be identified in a laminated fiber composite panel, where a crack like feature propagates severing all laminae of the laminate. Then, by unit area of crack surface, we mean unit area of completely failed (both matrix and fiber failure) surface along the crack like diffused zone of damage. The usage of the term ``fracture toughness'' is resisted, to avoid confusion with the traditional LEFM approach, where it is often regarded as a material property.} with increase in specimen size of single edge notch three point bend (SETB) specimen, subjected to primarily Mode-I crack tip conditions. The material used in all the experiments herein is a carbon fiber/epoxy $[-45/0/+45/90]_{6s}$ laminated fiber reinforced composite with a volume fraction of 0.55, and whose lamina and laminate properties are given in Table ~\ref{table:Material}.\\

The SETB configuration used in this study is shown in Figure ~\ref{fig:SENBEXP}. SETB specimens were cut from the composite panels using water jet facility. The notch was introduced and a knife edge was used to introduce a sharp starter crack. The specimens were supported on rubber rollers both at the loading and support points to minimize any local inelastic deformation. The specimens were loaded on a specially designed loading frame with anti buckling guide rods that prevents out of plane movement of the specimens. The specimens were loaded at a rate of $0.01mm/sec$ using hydraulically operated MTS testing machine and were loaded until failure. Load was measured by a load cell and the load point displacement was measured in between the top and bottom loading rollers using an LVDT. Five specimen sizes with geometrically scaled planar geometry and fixed thickness were considered.  Multiple specimens of each size were tested to significantly capture the failure response envelope. The load - load point displacement($P\Delta$) responses of these specimens are shown in Figure ~\ref{fig:SENBResults}. The apparent in-plane fracture resistance (hereafter referred to as the``in-plane fracture resistance'', or simply as ``fracture resistance'') of each specimen is determined by inverse modeling, i.e, by determining the value of fracture resistance input required in the VMCM model (Section ~\ref{micro}), such that the experimental and numerical $P\Delta$ curves and the corresponding instantaneous crack lengths nearly match. The multiple tests for each SETB size show some scatter in the $P\Delta$ response. Hence, an averaged fracture resistance, $\mathscr{R}_{avg}$, of each specimen is calculated by averaging the individual in-plane fracture resistance values obtained by inverse-modeling from each of the $P\Delta$ curves of that specimen.\footnote{It is interesting to note that in this material, the value of fracture resistance obtained by inverse modeling nearly matches the value obtained by normalizing the total area under the $P\Delta$ curve by the area (thickness times total crack length) of the crack surface formed.}\footnote{One may suggest the R-curve approach of LEFM as a better alternative. However the failure response of this material is stochastic, and lacks post-peak $P\Delta$ response repeatability, which is important for obtaining consistent R-curves. Averaging R-curves across specimens of each size is possible, but it is of little advantage over $\mathscr{R}_{avg}$. Furthermore, the total energy released may include dissipation due to other mechanisms, like crack tip inelasticity, but for this class of materials, the crack formation energy, which includes fiber pullout, fiber breaking and matrix cracking, is the dominant component of the total energy dissipated.} As will be demonstrated and discussed in Section ~\ref{simulations}, $\mathscr{R}_{avg}$ captures the correct failure response, and it is also much easier to determine. \\

Table ~\ref{table:Scaling} summarizes the observed scaling in the $P\Delta$ response, and hence in the value of $\mathscr{R}_{avg}$. These quantities $P^{\ast}$, $\Delta^{\ast}$ and $\mathscr{R}_{avg}^{\ast}$ are fixed reference values, of which $\mathscr{R}_{avg}^{\ast}$ is the value of fracture resistance obtained in similar tests on standard Compact Tension (CT) specimens (Figure ~\ref{fig:CTSEXP}). The scaling observed in $\mathscr{R}_{avg}$ is very significant, because in all the specimen sizes considered, it has been experimentally observed that there is formation of a Full Bridging Zone (FBZ) (Figure ~\ref{fig:SENBEXP2}), and the length of FBZ scales up with specimen geometry.  FBZ is assumed to have formed when fibers at the location of the initial crack tip have failed/completely pulled out, leading to zero tractions across the crack faces at the initial crack tip (\citet{Cox1991}). FBZ formation is also referred to as the formation of a ``stable process zone''. When the fibers at the initial crack tip have not yet failed, the bridging zone is still evolving, and herein we refer to this as the partial bridging zone.\\

The formation of a FBZ in all specimen sizes considered and the observed scaling in its length are contrary to the usual expectation, wherein, there is no FBZ formation prior to complete failure in smaller specimens, and when FBZ formation does occur in large enough specimens, the FBZ length is a material property. This assumption of a constant length FBZ formation can lead to incorrect predictions of strength and reliability, because in some classes of materials, like the one being considered here, a FBZ formation can be misinterpreted as the formation of a ``converged process zone'', whereas, in reality, larger size specimens might still lead to longer process zones, and hence greater FBZ length. Here, the phrase ``converged process zone'' refers to a process zone that does not increase in length if the specimen size is increased beyond a certain size. That is, the converged process zone length is invariant with respect to further increase in specimen dimensions. In order to substantiate and understand the mechanics behind such scaling behavior and FBZ formation, VMCM based simulations were conducted. Before proceeding to the numerical simulations, the VMCM framework is briefly presented in the following section.\\

\section{Mathematical Formulation}
\label{VMCM}

The standard weak form of the balance of linear momentum over the domain $\Omega$ (Figure ~\ref{fig:fine_body}) is given by,
%
\begin{equation}
\label{eqn:weak_form}
   \int_{\Omega} \nabla^{s} \boldsymbol{\it w} \colon
   \boldsymbol{\sigma} \ \mbox{dV} = \int_{\Omega} \boldsymbol{\it w}
   \cdotp \boldsymbol{\it b}
   \ \mbox{dV} + \int_{\partial \Omega_{t}}
   \boldsymbol{\it w} \cdotp \boldsymbol{\it T} \ \mbox{dS}.
\end{equation}

\noindent{where $\boldsymbol{\sigma}$ is the stress, $\boldsymbol{\it w}$ is an admissible displacement
variation, $\nabla^{s}\boldsymbol{\it w}$ is the symmetric gradient of the variation, $\boldsymbol{T}$ is the external traction and $\boldsymbol{b}$ is the body force.\\

In the standard finite element formulation of continuum mechanics, the displacements are at least $C^{0}$ continuous. But in a wide class of problems (shear banding, fiber kink banding, transverse crack formation, delamination initiation are some examples), the displacement field can be discontinuous. In such cases, the displacement field can be decomposed into continuous coarse scale and discontinuous fine scale components (Figure ~\ref{fig:Scale_Seperation}). Such a decomposition is also imposed upon the displacement variation, $\boldsymbol{\it w}$. The decomposition is made precise by requiring that the fine scales, $\boldsymbol{\it u'}$ and $\boldsymbol{\it w'}$, vanish outside of some region $\Omega^{\prime}$, which will be referred to as the microstructural or fine scale subdomain. This decomposition is written as,

\begin{equation}
\label{eqn:discon_scale}
    \boldsymbol{\it u} = \underbrace{\bar{\boldsymbol{\it u}}}_{\mbox{coarse scale}} + \underbrace{\boldsymbol{\it u'}}_{\mbox{fine scale}}
\end{equation}

\noindent{The corresponding scale separation in the displacement variation is given by,}

\begin{equation}
\label{eqn:discon_scale_W}
    \boldsymbol{\it w} = \underbrace{\bar{\boldsymbol{\it w}}}_{\mbox{coarse scale}} + \underbrace{\boldsymbol{\it w'}}_{\mbox{fine scale}}
\end{equation}

\begin{equation*}
\label{eqn:discon_scale_U_W}
\boldsymbol{\it u'}, \boldsymbol{\it w'} \varepsilon {\it S'} = \{ \boldsymbol{\it v'}|\boldsymbol{\it v'}=0  \hspace{2 mm} on  \hspace{2 mm} \Omega \setminus int(\Omega^{\prime}) \}  
\end{equation*}

Substituting the above decomposition into (\ref{eqn:weak_form}), and using standard arguments, the weak form
can be split into two separate weak forms. One, involving the coarse scale variation, $\bar{\boldsymbol{\it w}}$,
and the other, involving only the fine scale variation, $\boldsymbol{\it w'}$.

\begin{equation}
\label{eqn:weak_form_1}
   \int_{\Omega} \nabla^{s} \bar{\boldsymbol{\it w}} \colon
   \boldsymbol{\sigma} \ \mbox{dV} = \int_{\Omega} \bar{\boldsymbol{\it w}}
   \cdotp \boldsymbol{\it b}
   \ \mbox{dV} + \int_{\partial \Omega_{t}}
   \bar{\boldsymbol{\it w}} \cdotp \boldsymbol{\it T} \ \mbox{dS}.
\end{equation}

\begin{equation}
\label{eqn:weak_form_2}
   \int_{\Omega'} \nabla^{s} \boldsymbol{\it w'} \colon
   \boldsymbol{\sigma} \ \mbox{dV} = \int_{\Omega'} \boldsymbol{\it w'}
   \cdotp \boldsymbol{\it b}
   \ \mbox{dV} + \int_{\partial \Omega'_{t}}
   \boldsymbol{\it w'} \cdotp \boldsymbol{\it T} \ \mbox{dS}.
\end{equation}

This procedure results in the fine scale weak form (\ref{eqn:weak_form_2}), defined only
over $\Omega^{\prime}$ (Figure ~\ref{fig:fine_body}). This result is crucial since it lends
itself naturally to the application of desired micromechanical descriptions restricted to the
microstructural region, $\Omega'$, and not the entire domain $\Omega$. The scale separation in $\boldsymbol{\it u}$ is
contained in $\sigma=\mathcal{C} \colon (\nabla^{s} \bar{\boldsymbol{\it u}} + \nabla^{s} \boldsymbol{\it u'})$,
where $\mathcal{C}$ is the elastic stiffness tensor.\\

We wish to use an appropriate micromechanical law by which the fine scale solution, $\boldsymbol{\it u'}$, can be
expressed in terms of $\bar{\boldsymbol{\it u}}$ and other fields in the problem. Below, we will show how such a
micromechanical law can be embedded into the formulation using the weak form (\ref{eqn:weak_form_2}). The final
step involves elimination of the fine scale displacement, $\boldsymbol{\it u'}$, from the problem by substituting
its relation to $\bar{\boldsymbol{\it u}}$ in the coarse scale weak form (\ref{eqn:weak_form_1}). Thus, the fine
scale solution does not appear explicitly; however, its effect is fully embedded in the resultant modified weak form. \\

We choose $\Omega^{\prime}$ to contain the crack surface $\Gamma$ on which $\boldsymbol{\it u'}$ is discontinuous.
Invoking standard variational arguments, the weak form of the fine scale problem can be reduced to the following
statement of traction continuity (\citet{Garikipati2002}):

\begin{equation}
\label{eqn:traction_continuity}
   {\llbracket \boldsymbol{\sigma \it n} \rrbracket}_{\Gamma}=0
\end{equation}

\noindent{where $\llbracket . \rrbracket$ is the discontinuity in the quantity and $\boldsymbol{\it n}$ is the normal to the crack surface, $\Gamma$. Writing the traction on $\Gamma$ in terms of components $T_{n}$ and $T_{m}$ along $\boldsymbol{\it n}$ and $\boldsymbol{\it m}$ respectively (Figure ~\ref{fig:fine_body}), the traction continuity condition can be expressed as,}

\begin{equation}
\label{eqn:traction_continuity_2}
   T_{n} \boldsymbol{\it n} + T_{m} \boldsymbol{\it m} =  \boldsymbol{\sigma \it n} |_{\Gamma^{-}}
\end{equation}

The traction $\boldsymbol{\sigma \it n} |_{\Gamma^{-}}$, is determined by the macromechanical continuum formulation.  The
evolution of $T_{n}$ and $T_{m}$  is governed by the micromechanical surface law (Section ~\ref{simulations}(A)). This law, which emerges at a {\it finite} value of traction, is related to the displacement discontinuity which is the separation between the surfaces. There are two traction laws across a planar (2D, line) surface and the displacement discontinuity $\llbracket \boldsymbol{u} \rrbracket$ can be expressed in terms of the normal opening, $\llbracket \boldsymbol{u} \rrbracket . \boldsymbol{\it n}$, and tangential slip, $\llbracket \boldsymbol{u} \rrbracket . \boldsymbol{\it m}$, across $\Gamma$. \\

We now consider a specific functional form for the micromechanical model, emerging at a non-vanishing traction,

\begin{eqnarray}
\label{eqn:band_soft_1}
   T_{n} = T_{n_{0}} - \mathcal{H}_{n} \llbracket \boldsymbol{u} \rrbracket . \boldsymbol{\it n} , && T_{m} = T_{m_{0}} - \mathcal{H}_{m} \llbracket \boldsymbol{u} \rrbracket . \boldsymbol{\it m}
\end{eqnarray}

\noindent{where $T_{n_{0}}$ and $T_{m_{0}}$ are the maximum values of $T_{n}$ and $T_{m}$ admissible on ${\Gamma}$ (Figure ~\ref{fig:fine_body}), $\llbracket \boldsymbol{u} \rrbracket . \boldsymbol{\it n}>0$ is the normal jump (Mode-I type crack opening) and $\llbracket \boldsymbol{u} \rrbracket . \boldsymbol{\it m}$ is the tangential slip (Mode-II type crack face slip) along the elemental crack face, $\mathcal{H}_{n}$ and $\mathcal{H}_{m}$ are the softening moduli for the Mode-I and Mode-II crack opening evolution, respectively. Consistency between the micromechanical law and the macromechanical continuum description is enforced by (\ref{eqn:traction_continuity_2}) via (\ref{eqn:band_soft_1}).}\\

Substituting (\ref{eqn:band_soft_1}) in (\ref{eqn:traction_continuity_2}) and dispensing with the explicit indication of $\boldsymbol{\sigma \it n} |_{\Gamma^{-}}$,

\begin{eqnarray}
\label{eqn:band_soft_2}
  (T_{n_{0}} - \mathcal{H}_{n} (\llbracket \boldsymbol{u} \rrbracket.\boldsymbol{\it n}))\boldsymbol{\it n} +
  (T_{m_{0}} - \mathcal{H}_{m} (\llbracket \boldsymbol{u} \rrbracket.\boldsymbol{\it m}))\boldsymbol{\it m} - \boldsymbol{\sigma \it n} = 0
\end{eqnarray}

Expanding (\ref{eqn:band_soft_2}) up to first order terms, in order to solve for $\boldsymbol{\it u'}$:

\begin{eqnarray}
\label{eqn:band_soft_3}
  (T_{n_{0}} - \mathcal{H}_{n} (\llbracket \boldsymbol{u} \rrbracket.\boldsymbol{\it n}))\boldsymbol{\it n} +
  (T_{m_{0}} - \mathcal{H}_{m} (\llbracket \boldsymbol{u} \rrbracket.\boldsymbol{\it m}))\boldsymbol{\it m} - \boldsymbol{\sigma \it n} \nonumber \\
  - \mathcal{H}_{n}(\delta \llbracket \boldsymbol{u} \rrbracket. \boldsymbol{\it n}) \boldsymbol{\it n}
  - \mathcal{H}_{m}(\delta \llbracket \boldsymbol{u} \rrbracket. \boldsymbol{\it m}) \boldsymbol{\it m} \nonumber \\
  - (\mathcal{C} \colon (\nabla \delta \bar{\boldsymbol{u}} + \nabla \delta \boldsymbol{u'})) \boldsymbol{\it n} &=& 0
\end{eqnarray}

\noindent{where the first line in (\ref{eqn:band_soft_3}) represents a zeroth-order approximation to (\ref{eqn:band_soft_2}), and the remaining terms are the first order corrections. Using $\boldsymbol{u'}=\llbracket \boldsymbol{u} \rrbracket \it C_{\Gamma}$, where $\it C_{\Gamma}$ is the fine scale interpolation (Figure ~\ref{fig:discontinousShapeFunction}), converts (\ref{eqn:band_soft_3}) into a linear equation in $\delta \llbracket \boldsymbol{u} \rrbracket$ which can be solved, and then the incremental fine scale field is obtained from $\delta \boldsymbol{u'} = \delta \llbracket \boldsymbol{u} \rrbracket \it{C_{\Gamma}}$. Formally, it is represented as,}

\begin{equation}
\label{eqn:band_soft_4}
  \delta {\boldsymbol{u'}}=F[\bar{\boldsymbol{u}}, \boldsymbol{\sigma}, T_{n}, T_{m}, \xi_{n}, \xi_{m}]
\end{equation}

Extending the incremental formulation to $\boldsymbol{\sigma}$, which in a general nonlinear problem can be expanded up to first order as $\boldsymbol{\sigma}=\boldsymbol{\sigma}^{(0)}+ \mathcal{C} \colon (\nabla \delta \bar{\boldsymbol{u}} + \nabla \delta \boldsymbol{u'})$, where $\boldsymbol{\sigma}^{(0)}$ is the converged value of $\boldsymbol{\sigma}$ in the previous solution increment, and substituting $\boldsymbol{u'}$ in (\ref{eqn:weak_form_1}), we obtain the coarse field weak form which is independent of the fine scale displacement $\boldsymbol{u'}$. On solving for $\delta \bar{\boldsymbol{u}}$, the incremental fine scale field $\delta {\boldsymbol{u'}}$ can be recovered via (\ref{eqn:band_soft_4}). Iterations are to be performed:
${\bar{\boldsymbol{u}}}^{(i+1)}={\bar{\boldsymbol{u}}}^{(i)} + \delta \bar{\boldsymbol{u}}$, ${{\boldsymbol{u'}}}^{(i+1)}={{\boldsymbol{u'}}}^{(i)} + \delta {\boldsymbol{u'}}$, until a converged solution is
obtained. From (\ref{eqn:weak_form_1}),(\ref{eqn:band_soft_3}) and (\ref{eqn:band_soft_4}), it should be clear that the VMCM method results in an embedding of the micromechanical surface law into the coarse scale weak formulation. Interested readers are referred to \citet{Garikipati2002} for a more detailed discussion of the numerical framework.\\

Having briefly presented the numerical formulation, we direct our attention to simulations, and the
mechanisms involved in the in-plane fracture of laminated fiber composites.

\section{Numerical Simulations}
\label{simulations}

\subsection{Micromechanical surface law}
\label{micro}

The micromechanics of crack propagation is embedded into the macroscopic formulation of the VMCM by enforcing the linear traction evolution relations (\ref{eqn:band_soft_2}). Due to the stochastic behavior observed in experiments and nonlinearity of response (Section ~\ref{discussion}), it is difficult to obtain precise functional forms for $\mathcal{H}_{n}$ and $\mathcal{H}_{m}$. The accepted methodology in the cohesive zone community (\citet{Hillerborg1976}, \citet{Dugdale1960}, \citet{Barenblatt1962}) is to assume a suitable functional form of the traction evolution response, referred to as the traction separation law, anchored by experimentally determined values of $T_{n_{0}}$, Mode-I $\mathscr{R}_{avg}$ and $T_{m_{0}}$, Mode-II $\mathscr{R}_{avg}$. In the current study, a linear traction separation law (Figure ~\ref{fig:LinSoft}) is assumed for the evolution of $T_{n}$.\footnote{In order to avoid additional physical complexities arising from mode mixity conditions only test cases undergoing pure Mode-I evolution are considered in the current study, and hence there is no evolution of $T_{m}$. However, the VMCM framework is general and works just as well for problems involving curved crack propagation and mixed-mode evolution (\citet{Garikipati2002}).} There is no theoretical or physical rationale in assuming a linear traction separation law as against any other functional form\footnote{For this class of materials the simulations are more sensitive to the values of $T_{n_{0}}$ and $\mathscr{R}_{avg}$, than the shape of the tractions separation curve.}. The actual shape of the traction separation law is still a topic of active research (\citet{Brocks2003}) and some methods have been devised to measure the law for a limited class of configurations (\citet{Stigh2004}, \citet{Sorensen1998}).\\

The value of $T_{n_{0}}$ is obtained by experiments on double notched tension specimens (Figure ~\ref{fig:TENEXP}). This configuration was selected because the stress state across the entire crack face is almost uniform and specimen failure is instantaneous. Thus, the critical load divided by the total crack area gives a fairly accurate estimate of the critical traction across the crack faces. It was observed that this value was independent of the specimen size. The $\mathscr{R}_{avg}$ values are given in Table (\ref{table:Scaling}). With these two inputs, the five SETB specimen sizes given in Figure ~\ref{fig:SENBEXP} are simulated using the VMCM. Further, the crack is propagated perpendicular to the maximum principal stress direction.\\


\subsection{Simulations}
\label{simulation}

The criterion selected in this work for validating the VMCM methodology is the comparison of VMCM and experimental $P\Delta$ responses of SETB specimens (Figure~\ref{fig:SENBEXP}). This criterion is chosen because the $P\Delta$ response reflects the macroscopic response of the structure to external loads, and often in structural design the peak load is the value of primary interest. A representative finite element simulation mesh is shown in Figure ~\ref{fig:senbMesh}. Figure ~\ref{fig:PDFixedR} shows the $P\Delta$ curves extracted from VMCM simulations of SETB specimens with a constant, $\mathscr{R}_{avg}=\mathscr{R}_{avg}^{\ast}$, input for all sizes considered. As seen in the figure, the load displacement response is significantly captured for smaller specimens (because $\mathscr{R}_{avg}^{\ast}$ is close to $\mathscr{R}_{avg}$ value for Size-1 and Size-2 specimens), but the peak load is severely under-predicted for larger specimens. This is expected, as the experimental values of $\mathscr{R}_{avg}$ for large specimens are much higher then the input value $\mathscr{R}_{avg}^{\ast}$ (Table~\ref{table:Scaling}). Now the simulations are conducted with the experimentally measured $\mathscr{R}_{avg}$ values given in Table~\ref{table:Scaling}, and the results are plotted in Figure ~\ref{fig:PDScaledR}. \\

As seen from Figure ~\ref{fig:PDScaledR}, the VMCM simulations accurately reproduce the macroscopic response of the SETB specimens when appropriate $\mathscr{R}_{avg}$ values are used as input.\footnote{The small pre-peak stiffness difference in the experimental curves of some specimens is due to some amount of material crushing occurring at the loading points, and the post-peak variations are an artifact of the stochastic nature of failure in these materials.} This demonstrates that: (1) The VMCM methodology has the ability to numerically simulate progressive damage propagation, and the mechanics of bridged crack evolution. (2) In spite of a multitude of failure mechanisms operating simultaneously, leading to a very complex evolution of fracture resistance, the single valued estimate, $\mathscr{R}_{avg}$, of the fracture resistance is appropriate for numerical simulations, at least in this class of materials.\footnote{Even though the linear traction separation law simplifies a set of complex failure mechanisms, the dissipated energy is well-represented. This quantity appears to control the fit with experimental $P\Delta$ curves.} \\

Having presented the relevance of VMCM and $\mathscr{R}_{avg}$, we now turn to the actual mechanics of the observed scaling behavior: Why does the fracture resistance (and hence $\mathscr{R}_{avg}$) increase with specimen size?. As a start, we use VMCM simulations to study the effect of high fracture resistance on the $P\Delta$ response of this material model. Figure ~\ref{fig:PDRangeR} shows the effect of varying $\mathscr{R}_{avg}$ on the peak $P\Delta$ response of the specimens. As seen, increase in $\mathscr{R}_{avg}$ leads to plateauing of the peak response. The peak load increases only slightly, but the tendency for softening past the peak is suppressed. One inference from this observation is that in this class of materials, high fracture resistance has only limited influence in increasing the peak load carrying ability of the specimens. Beyond a certain range of fracture resistance, the specimen geometry starts to play a significant role. This is evident in Figure ~\ref{fig:bridgingSENB1}, where the bridging zone formation and movement is depicted. As seen in the figure, after the formation of a FBZ, the zone propagates for some distance with almost a constant length. But as the FBZ approaches the end of the crack path, the compressive stresses encountered by the growing crack, caused due to bending of the SETB specimen about the loading point, inhibit further FBZ propagation, and hence the FBZ length decreases.\footnote{The maximum FBZ length corresponds to the peak load. Any further progress of the FBZ has to be accompanied by a reduction in the FBZ length, and hence drop in $P$, to satisfy force and moment equilibrium of the specimen. Any further reduction in FBZ length due to compressive stresses ahead of the crack tip, is in addition to the reduction of the FBZ length necessitated due to equilibrium conditions.}\\

Based on the simulation results depicting the influence of $\mathscr{R}_{avg}$ on the $P\Delta$ response (Figure ~\ref{fig:PDFixedR}, ~\ref{fig:PDRangeR}) and the FBZ formation (Figure ~\ref{fig:bridgingSENB1}), we present the following assessment of the failure behavior of this class of materials. It appears that this class of material is capable of achieving a very large bridging zone length, when there is no geometric constraint (both, ahead of the crack tip, and in the crack wake undergoing fiber pullout) on its natural bridging zone formation tendency. The corresponding maximum ``full bridging zone'' length is referred to as the ``free bridging zone''(FrBZ) length, which may be considered a material property. In this FrBZ, a multitude of mechanisms (fiber failure, debounding, pullout) are active over large lengths, leading to a high fracture resistance. But in specimens with sizes comparable to these large lengths, the FrBZ formation mechanisms are inhibited and the specimen fails to manifest the FrBZ, but instead forms a smaller FBZ. The smaller the specimen size, larger the inhibition to free bridging zone formation, and thereby, smaller the corresponding FBZ length manifested. This reduced FBZ length leads to lower fracture resistance, and hence lower peak load. \\

\section{Discussion}
\label{discussion}

The material in consideration is composed of a quasi-brittle matrix and reinforcing carbon fibers. For this material, experiments show at first a matrix crack forming, leading to enhanced loading on the adjoining fibers, which suffer greater strain until they reach their failure stress. Ultimate failure occurs at weak points either at the matrix crack interface or inside the surrounding matrix. Fiber failure at the matrix crack interface leads to complete loss of load bearing capacity at the respective fiber. If the fiber fails at a weak point inside the intact matrix, it leads to fiber pull out, which involves debonding failure and frictional resistance to the fiber movement (\citet{Cooper1970}). As stated earlier, the process zone over which these mechanisms occur is termed the FBZ. These additional dissipation mechanisms contribute to an increase in fracture resistance until the fiber is completely pulled out of the matrix. These mechanisms are schematically depicted in Figure ~\ref{fig:cracking1}. Theoretical models have been developed to study fiber failure and pullout, and its effect on fracture resistance (\citet{Cooper1970}, \citet{Kelly1971}, \citet{Kelly1973}, \citet{Cox1991}). These models are useful in gaining insight to the individual contributions of various constituent mechanisms towards fracture resistance, but in a real material the observed fracture resistance is a combination of these processes. Furthermore, the actual crack path formation is significantly stochastic, and it is prohibitively difficult, if not impossible, to delineate the contribution of each process. Thus, a macroscopic representation of fracture resistance, as done here, provides a convenient means to model through the thickness crack growth in this class of materials.\\

In Section ~\ref{simulation}, the effect of varying $\mathscr{R}_{avg}$ on the peak load of the specimen was briefly discussed. Figure ~\ref{fig:PDRangeR} depicts the limited influence of $\mathscr{R}_{avg}$ in increasing the peak load of the specimen. As evident from the figure, beyond a certain range, increase in $\mathscr{R}_{avg}$ causes a plateauing of the $P\Delta$ response for the following reasons: \\
(1) Another parameter, $T_{n_{0}}$ (Equation ~\ref{eqn:band_soft_1}, Figure ~\ref{fig:LinSoft}), also influences the peak load magnitude. For any given FBZ length, the load carrying ability of the FBZ may be determined by adding the instantaneous load carried by all the individual fibers making up the FBZ. But the instantaneous load ($T_{n}$) carried by each fiber is dictated by Equation ~\ref{eqn:band_soft_1}, and thus $T_{n_{0}}$ influences the peak load of the specimen.\\
(2) Though $\mathscr{R}_{avg}$ and $T_{n_{0}}$ determine the maximum load carrying ability of the material, the actual evolution of the $P\Delta$ response has to be in accordance with the force and moment equilibrium conditions.\\

It is instructive to place the results that have been presented in the context of energetic size effect laws that have been developed by \citet{Bazant1970}. When geometrically scaled specimens are stressed to failure, it has been shown that many types of quasi-brittle materials obey a Type 2 size effect law that is described by \citet{Bazant1970}. As shown there, performing a regression analysis of just the maximum loads for the SETB specimens (Figure ~\ref{fig:SENBResults}) leads to estimates of FrBZ length and $\mathscr{R}_{max}$ in accordance with the scaling law. Using this $\mathscr{R}_{max}$ value as input, the VMCM simulations predict the FrBZ length which is in good agreement with the scaling law estimate (FrBZ$_{VMCM}$/FrBZ$_{Scaling Law}$=1.08). This comparison adds further confidence and validity to the findings presented here through the use of the VMCM method.


\section{Conclusions}
\label{conclusion} In this paper, the VMCM method has been used to study in-plane fracture and damage growth in laminated fiber reinforced composite panels. Motivated by a series of experimental results, a novel approach that circumvents the drawbacks of traditional cohesive zone modeling approaches has been developed. The method has the advantage of being able to predict non-self similar crack growth along paths that do not need to coincide with element edges. Furthermore, knowledge of the intended crack growth path is not needed, and in fact is an outcome of the VMCM method. The results presented and the ensuing discussion demonstrate that:\\
(1) In-plane fracture of fiber reinforced laminates, unlike in metals, exhibits geometry and loading dependent fracture
resistance.\\ (2) Averaged fracture resistance ($\mathscr{R}_{avg}$) values reflect the macroscopic fracture resistance of these specimens.\\
(3) The existence of a ``Free Bridging Zone''(FrBZ) and maximum fracture resistance ($\mathscr{R}_{max}$) are characteristics of this material. The geometric inhibition of FrBZ evolution and associated convergence of $\mathscr{R}_{avg}$ to $\mathscr{R}_{max}$ lead to the observed scaling behavior. Not accounting for the existence of FrBZ and $\mathscr{R}_{max}$ posses the risk of incorrect predictions of strength and reliability, because in some classes of materials, like the one being considered here, the FBZ formation may wrongly signal the formation of a converged process zone (FrBZ).\\

\setcounter{secnumdepth}{0}

\section{Acknowledgments}
\label{acknowledgment}
This work has benefited from the financial support of Collier Research Corporation, through a NASA NRA award.
The interest and encouragement of Steven M. Arnold and Brett Bednarcyk of NASA Glenn Research Center is gratefully acknowledged.

\bibliographystyle{elsart-harv}
\bibliography{Rudraraju_IJSS}

\clearpage

\begin{table}
\caption{Lamina and laminate properties of carbon fiber/epoxy $[-45/0/+45/90]_{6s}$ laminated fiber reinforced composite.}
\centering
\begin{tabular}{|c|c|cc|c|}
	\hline 
Laminate & Lamina \\
	\hline \hline
$E_{xx}$: 51.5 GPa   & $E_{11}$: 141 GPa \\
$E_{yy}$: 51.5 GPa   & $E_{22}$: 6.7 GPa \\
$G_{xy}$: 19.4 GPa   & $G_{12}$: 3.2 GPa \\
$\nu_{xy}$: 0.32     & $\nu_{12}$: 0.33 \\
\hline
\end{tabular}
\label{table:Material}
\end{table}

\begin{table}
\caption{Scaling observed in the SETB specimen experiments.}
\centering
\begin{tabular}{|c|c|cc|c|}
	\hline 
Size & Geometry scaling & Peak load & Load point displacement & Fracture resistance \\ & (Figure ~\ref{fig:SENBEXP}) & $P/P^{\ast}$ & $\Delta / \Delta^{\ast}$ & $\mathscr{R}_{avg}/\mathscr{R}_{avg}^{\ast}$\\
	\hline \hline
1   & 1 & 0.27 & 0.1  & $1.08$\\
2   & 1.5 & 0.4 & 0.15 & $1.23$ \\
3   & 2 & 0.6 & 0.2 & $1.84$\\
4   & 3 & 0.81 & 0.28 & $2.46$\\
5   & 4 & 1.0 & 0.37  & $2.58$\\
	\hline
\end{tabular}
\label{table:Scaling}
\end{table}

\begin{figure}
  \begin{center}
  \psfrag{A}{20.3 mm}
  \psfrag{B}{17.7 mm}
  \psfrag{C}{7.6 mm}
  \psfrag{D}{3.8 mm}
  \psfrag{P}{$P, \Delta $}
  \includegraphics[width=4in,height=2in]{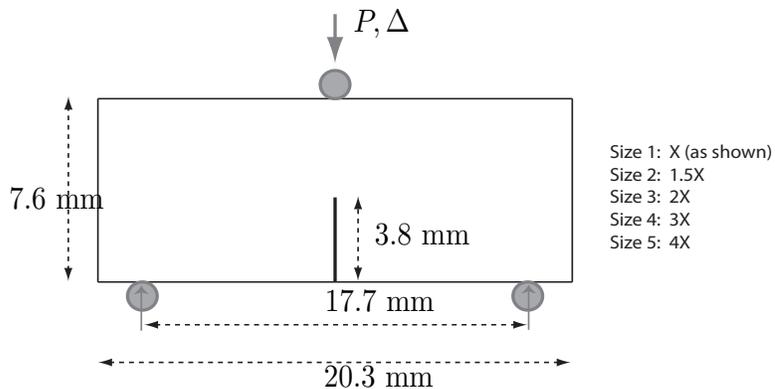}
  \end{center}
  \caption[SENB]{Single Edge Notch Bending(SETB) specimen configuration used for validating VMCM simulation results. Size 1 has the dimensions shown in figure, other sizes are scaled versions of this base size. All specimens have a nominal thickness of 6.35mm.}
  \label{fig:SENBEXP}
\end{figure}

\begin{figure}
  \begin{center}
  \psfrag{a}{Crack opening in Size 5 specimen}
  \psfrag{b}{Crack opening in Size 1 specimen}
  \psfrag{c}{Fiber bridging in Size 5 specimen}
  \psfrag{e}{2 mm}
  \psfrag{f}{1 mm}
  \includegraphics[width=6in,height=4in]{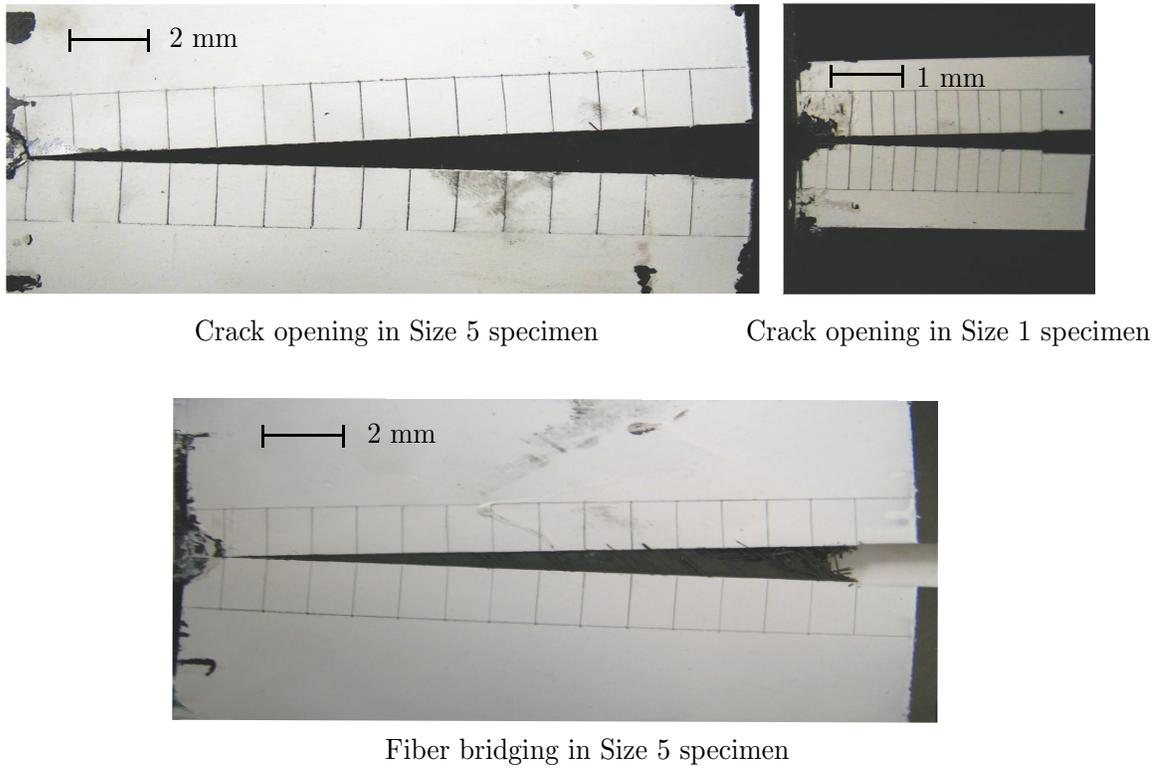}
  \end{center}
  \caption[SENB]{Comparison of crack tip opening in failed SETB size-1 and size-5 specimens.}
  \label{fig:SENBEXP2}
\end{figure}

\begin{figure}[t]
  \begin{center}
  \subfigure[][]{
  \psfrag{A}{$P, \Delta $}
  \psfrag{B}{\footnotesize{19 mm}}
  \psfrag{C}{\footnotesize{20.3 mm}}
  \psfrag{D}{7.6 mm}
  \psfrag{E}{9.5 mm}
  \includegraphics[width=2in,height=1.5in]{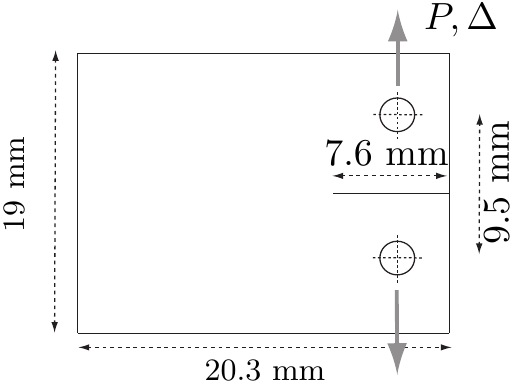}
  \label{fig:CTSEXP}}
  \hspace{.5in}
  \subfigure[][]{
  \psfrag{A}{6.3 mm}
  \psfrag{B}{4.3 mm}
  \psfrag{C}{1 mm}
  \psfrag{D}{$P$}
  \includegraphics[width=1in,height=1.5in]{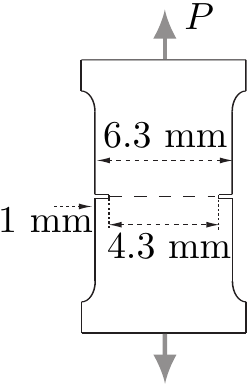}
  \label{fig:TENEXP}}
  \end{center}
  \caption[Specimen]{(a) Compact tension specimen (CTS) configuration used to obtain $\mathscr{R}_{avg}$ for Mode I crack propagation. (b) Double notch tension specimen configuration used to obtain critical cohesive stress value. All specimens have a nominal thickness of 6.35mm.}
  \label{fig:CTSTENEXP}
\end{figure}

\begin{figure}
  \begin{center}
  \psfrag{X}{Load point displacement ($\Delta / \Delta^{\ast}$)}
  \psfrag{Y}{Load ($P/P^{\ast}$)}
  \includegraphics[width=6in,height=2.5in]{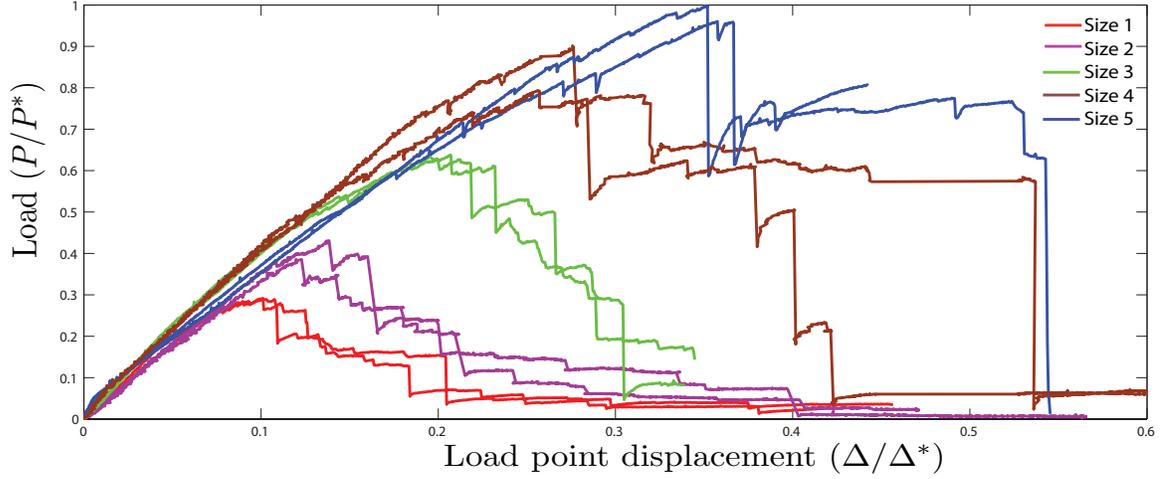}
  \end{center}
    \caption[senbresults]{Experimental Load ($P$) - Load point displacement ($\Delta $) curves obtained for various sizes of SETB specimens. Multiple specimens of each size were tested to capture the envelope of the failure response. $P^{\ast}$ and $\Delta^{\ast}$ are fixed reference values.}
    \label{fig:SENBResults}
\end{figure}


\begin{figure}[t]
	\begin{center}
	\psfrag{U}{$\boldsymbol{\it u} = \bar{\boldsymbol{\it u}} + \boldsymbol{\it u'}$}
	\psfrag{b}{$\bar{\boldsymbol{\it u}}$}
	\psfrag{d}{${\boldsymbol{\it u'}}$}
	\psfrag{J}{$\llbracket \boldsymbol{\it u} \rrbracket$}
	\includegraphics[width=4in, height=1.5in]{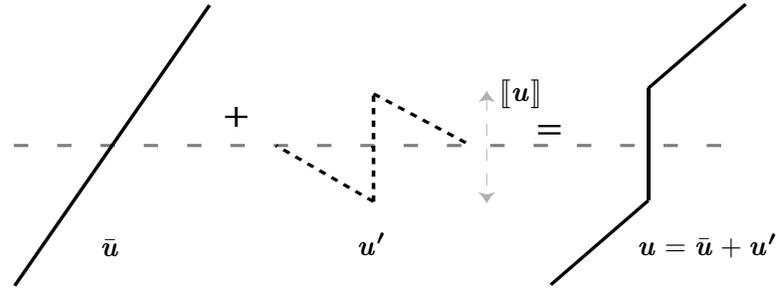}
	\end{center}
	\caption[]{Schematic of scale separation. $\bar{\boldsymbol{\it u}}$ is the coarse scale displacement field and  ${\boldsymbol{\it u'}}$ is the local fine scale enhancement.}
	\label{fig:Scale_Seperation}
\end{figure}

\begin{figure}[t]
  \begin{center}
  \psfrag{o}{$\Omega$}
  \psfrag{q}{}
  \psfrag{p}{}
  \psfrag{c}{$\Omega^{\prime}$}
  \psfrag{d}{}
  \psfrag{e}{}
  \psfrag{n}{$\boldsymbol{\it n}$}
  \psfrag{m}{$\boldsymbol{\it m}$}
  \psfrag{a}{}  
  \psfrag{r}{$\Gamma$}
  \psfrag{t}{}
  \psfrag{s}{}
  \includegraphics[width=2in,height=1.5in]{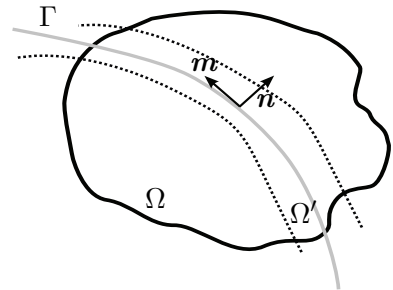}
  \end{center}
  \caption[]{Decomposition of continuum body into region where coarse scale and fine scale displacements are defined. $\Omega$ is the domain of the problem, $\Gamma$ is the displacement discontinuity (crack), $\Omega^{\prime}$ is the support for the displacement discontinuity and $\boldsymbol{\it n}$, $\boldsymbol{\it m}$ are the normal and tangent to $\Gamma$ respectively. }
  \label{fig:fine_body}
\end{figure}

\begin{figure}[t]
	\begin{center}
	\psfrag{U}{$\boldsymbol{\it C_{\Gamma}} = \boldsymbol{\it H_{\Gamma }} - \hat{\boldsymbol{\it N}}$}
	\psfrag{b}{$\hat{\boldsymbol{\it N}}$}
	\psfrag{d}{$\boldsymbol{\it H_{\Gamma }}$}
	\psfrag{h}{$\boldsymbol{\it h}$}
	\includegraphics[width=4in, height=1.5in]{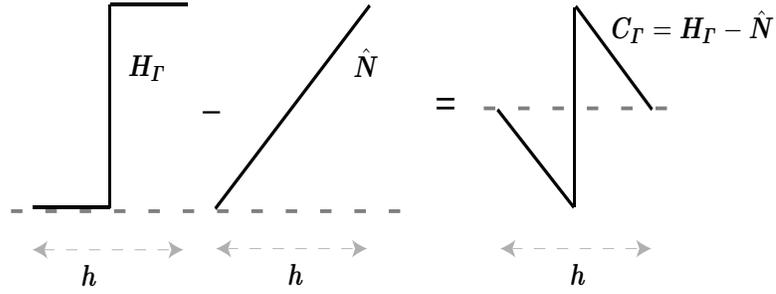}
	\end{center}
	\caption[]{Discontinuous shape function used to resolve the displacement jump shown in a one-dimensional setting. It is constructed by superimposing a discontinuous function, $\boldsymbol{\it H_{\Gamma }}$, on a regular polynomial, $\hat{\boldsymbol{\it N}}$. $\boldsymbol{\it h}$ is the element dimension.}
	\label{fig:discontinousShapeFunction}
\end{figure}

\begin{figure}[h]
  \begin{center}
  \psfrag{H}{$\mathcal{H}_{n}$}
  \psfrag{G}{$\mathscr{R}_{avg}$}
  \psfrag{x}{$\llbracket \boldsymbol{u} \rrbracket . \boldsymbol{\it n}$}
  \psfrag{p}{Normal crack opening}
  \psfrag{r}{$T_{n_{0}}$}
  \psfrag{s}{$ $}
  \psfrag{y}{$T_{n}$}
  \psfrag{q}[r][][1][90]{Resolved surface traction}
\includegraphics[width=2.00in]{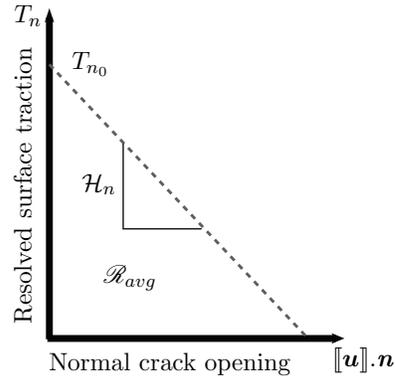}
\end{center}
   \caption[]{Linear micro-mechanical surface law for normal crack opening}
   \label{fig:LinSoft}
\end{figure}	

\begin{figure}[t]
	\begin{center}
	\small{
	\psfrag{A}{\textbf A}
	\psfrag{B}{\textbf B}
	\psfrag{C}{\textbf C}
	\psfrag{D}{\textbf A}
	\psfrag{E}{\textbf B}
	\psfrag{F}{\textbf C}
	\psfrag{L}{\textbf Load}
	\psfrag{P}{\textbf Displacement}
	\psfrag{X}{Matrix crack}
	\psfrag{Y}{Fiber bridging}
	\psfrag{Z}{FBZ formation}
	\psfrag{V}{FBZ propagation}
	\psfrag{W}{Region of failed fibers}
	\psfrag{O}{$\boldsymbol{\it \Omega}$}
	\psfrag{T}{$\boldsymbol{\it \Gamma}$}
	\includegraphics[width=5in, height=2in]{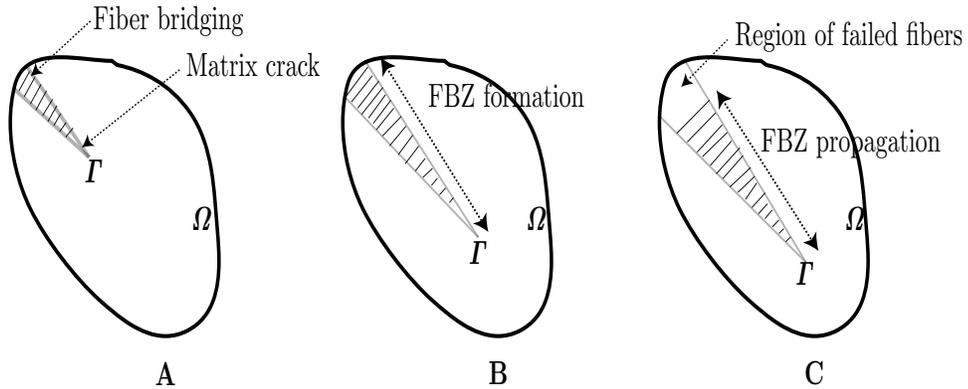}}
	\end{center}
	\caption[]{Stages involved in fiber composite cracking. A: Matrix cracking - Fiber bridging, B: FBZ formation, C: FBZ propagation.}
	\label{fig:cracking1}
\end{figure}

\begin{figure}[h]
  \begin{center}
  \subfigure[][Original mesh]{\label{fig:senbMeshZoom} \includegraphics[width=4.5in,height=2.5in]{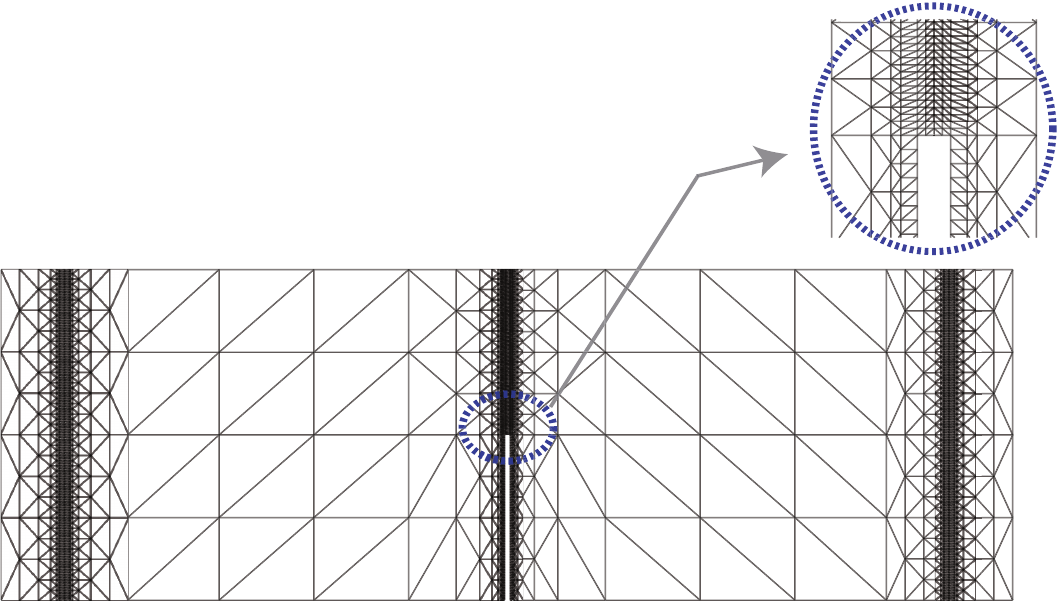}}
  \subfigure[][Deformed mesh: The red line represents the crack path]{\label{fig:senbMeshDeformed} \includegraphics[width=5in,height=2.5in]{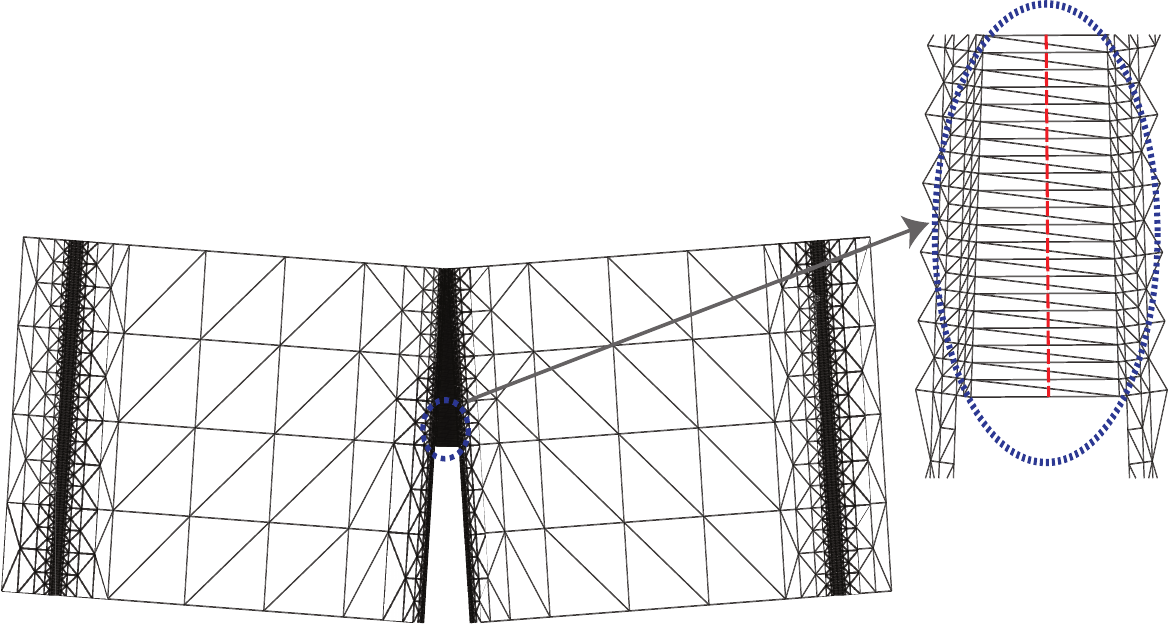}}
  \end{center}
  \caption[senbZoom]{SETB specimen mesh, with crack tip elements magnified in inset.}
  \label{fig:senbMesh}
\end{figure}

\begin{sidewaysfigure}[ht]
  \begin{center}
  \psfrag{A}{$\Delta $} \psfrag{B}{P} \psfrag{C}{$\Delta $} \psfrag{D}{P} 
  \psfrag{E}{$\Delta $} \psfrag{F}{P} \psfrag{G}{$\Delta $} \psfrag{H}{P}
  \psfrag{I}{$\Delta $} \psfrag{J}{P} \psfrag{K}{Size 1} \psfrag{L}{Size 2} 
  \psfrag{M}{Size 3} \psfrag{N}{Size 4} \psfrag{O}{Size 5}
  \includegraphics[width=8in,height=6in]{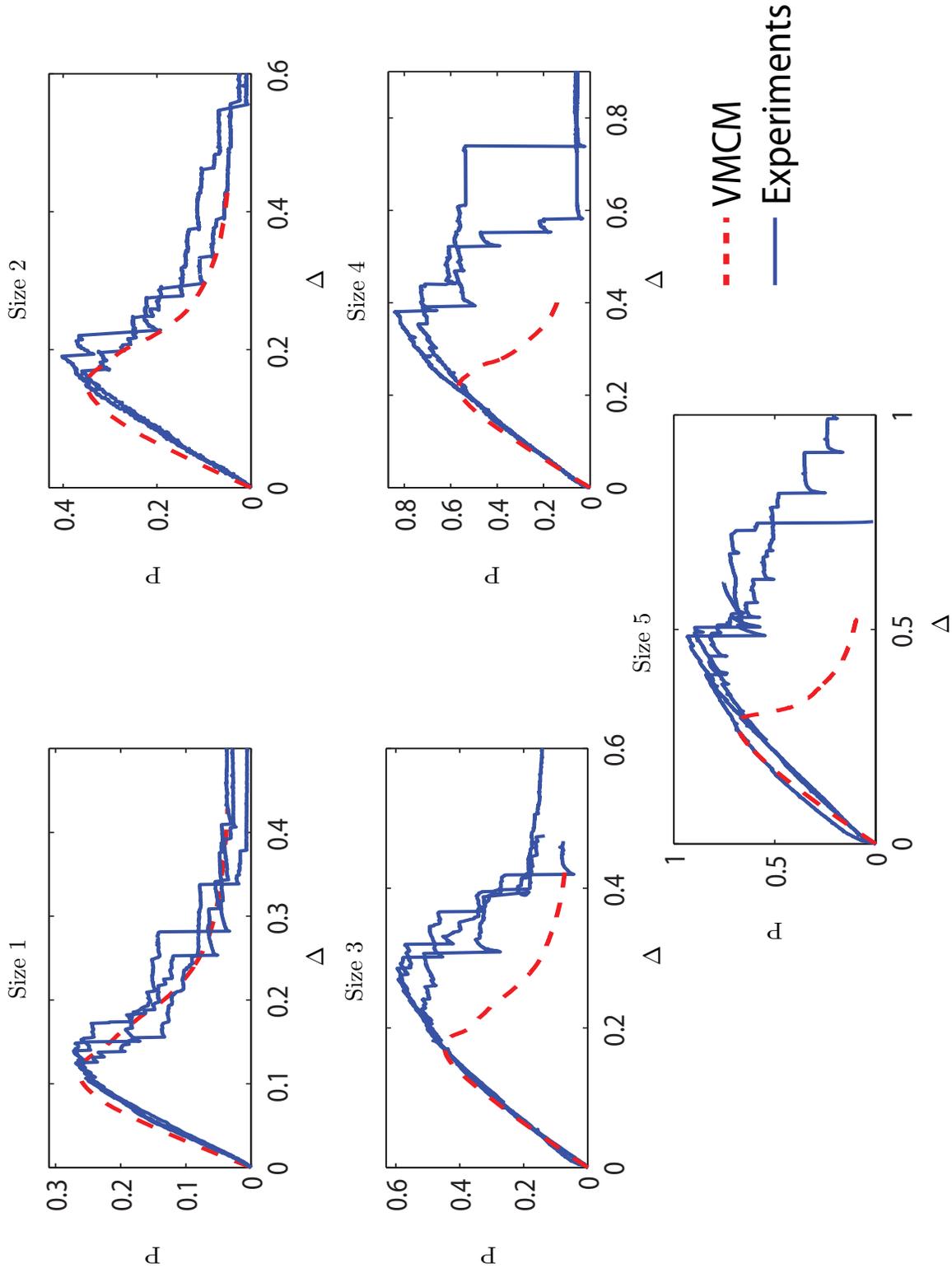}
  \end{center}
  \caption[]{Load-Displacement ($P\Delta$) response, extracted from VMCM simulations with a constant, $\mathscr{R}_{avg}=\mathscr{R}_{avg}^{\ast}$, input for Size 1-5 SETB specimens, compared with the experimental curves.  The $P$ and $\Delta$ values have been normalized with fixed reference values.}
  \label{fig:PDFixedR}
\end{sidewaysfigure}

\begin{sidewaysfigure}[ht]
  \begin{center}
  \psfrag{d}{$\Delta $} \psfrag{p}{P} \psfrag{V}{Size 1} \psfrag{W}{Size 2} 
  \psfrag{X}{Size 3} \psfrag{Y}{Size 4} \psfrag{Z}{Size 5} 
  \psfrag{A}{Experiments} \psfrag{B}{VMCM: $\mathscr{R}_{avg}= R_{L}, R_{H}$} \psfrag{C}{VMCM: $\mathscr{R}_{avg} = R_{M}$}
  \includegraphics[width=8in,height=6in]{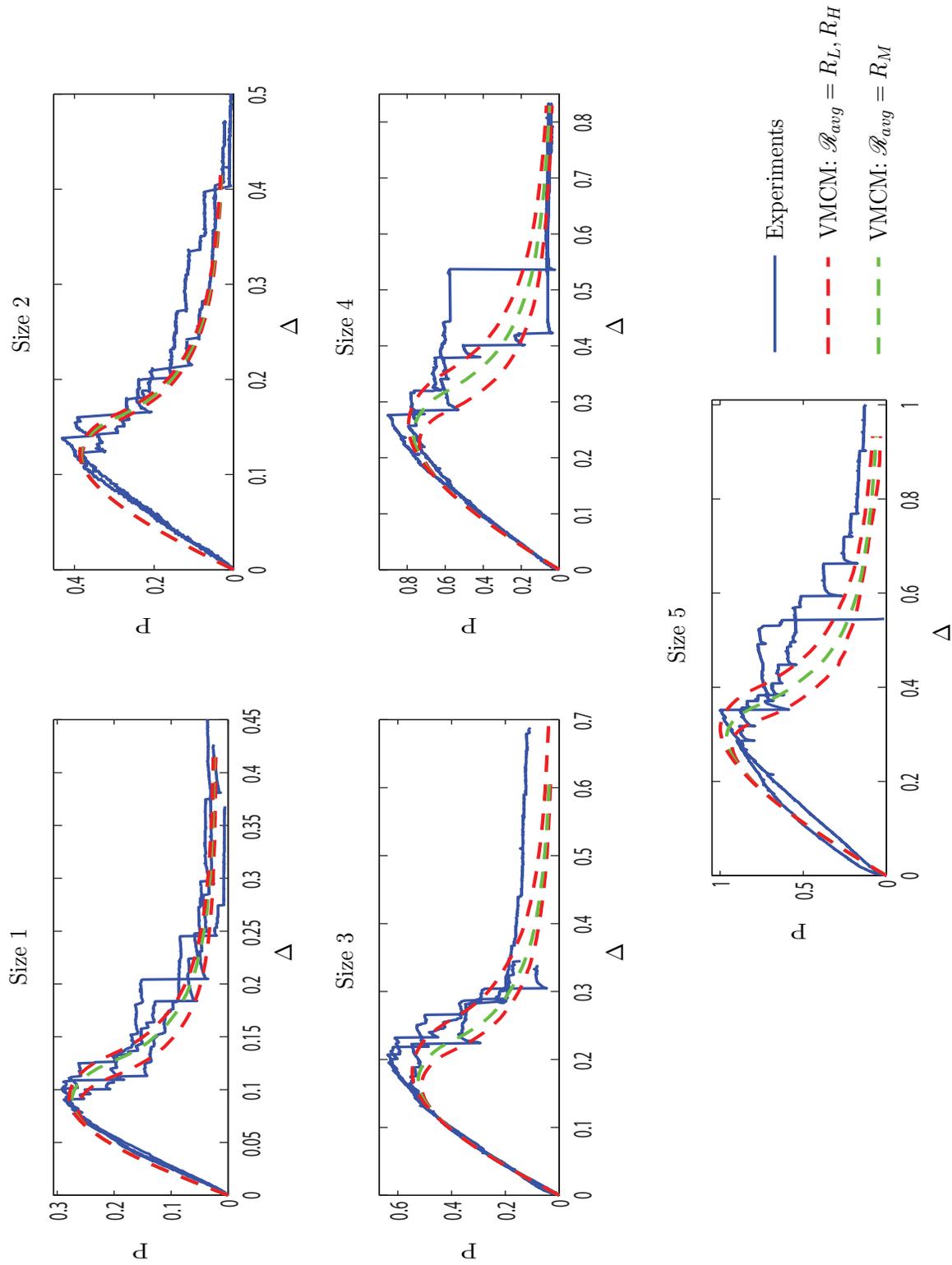}
  \end{center}
  \caption[]{Load-Displacement ($P\Delta$) response obtained from VMCM simulations of Size 1-5 SETB specimens with experimentally determined $\mathscr{R}_{avg}$ values (Table ~\ref{table:Scaling}), compared to their respective experimental curves. For a particular specimen size, $R_{L}$ and $R_{H}$ are the least and highest values of fracture resistance obtained from the multiple experimental $P\Delta$ curves, $R_{M}$ is the average of the fracture resistance of each of the multiple experimental $P\Delta$ curves. $R_{L}$ corresponds to the curve exhibiting least toughness and $R_{H}$ corresponds to the curve exhibiting the highest toughness.  The $P$ and $\Delta$ values have been normalized with fixed reference values.}
  \label{fig:PDScaledR}
\end{sidewaysfigure}

\begin{sidewaysfigure}[ht]
  \begin{center}
  \psfrag{d}{$\Delta $} \psfrag{p}{P} \psfrag{V}{Size 1} \psfrag{W}{Size 2} 
  \psfrag{X}{Size 3} \psfrag{Y}{Size 4} \psfrag{Z}{Size 5} 
  \psfrag{A}{$\mathscr{R}_{avg}=\mathscr{R}_{avg}^{\ast}$} \psfrag{B}{$\mathscr{R}_{avg}=2\mathscr{R}_{avg}^{\ast}$} \psfrag{C}{$\mathscr{R}_{avg}=3\mathscr{R}_{avg}^{\ast}$}  \psfrag{D}{$\mathscr{R}_{avg}=6\mathscr{R}_{avg}^{\ast}$} \psfrag{E}{$\mathscr{R}_{avg}: Experimental$} 
  \includegraphics[width=8in,height=6in]{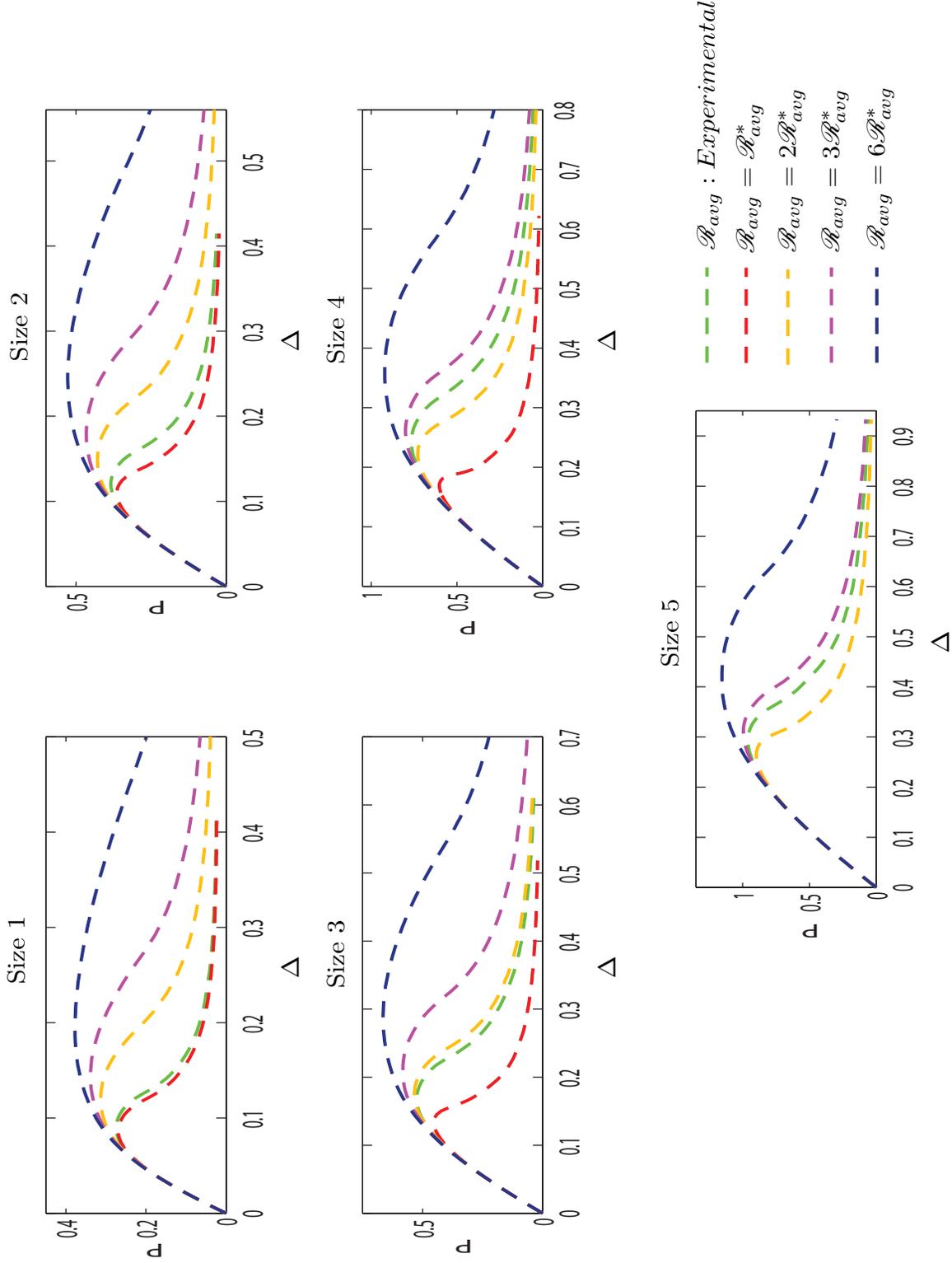}
  \end{center}
  \caption[]{Effect of increasing fracture resistance input on the Load-Displacement ($P\Delta$) response of VMCM simulations. $\mathscr{R}_{avg}^{\ast}$ is a constant reference value. Also plotted is the corresponding $P\Delta$ response obtained with the experimentally determined fracture resistance value. The $P$ and $\Delta$ values have been normalized with fixed reference values.}
  \label{fig:PDRangeR}
\end{sidewaysfigure}

\begin{sidewaysfigure}[ht]
 \psfrag{T}{Size 4 ($\mathscr{R}_{avg}=\mathscr{R}_{avg}^{\ast}$)}
 \psfrag{U}{Size 4 ($\mathscr{R}_{avg}=2.46\mathscr{R}_{avg}^{\ast}$)}
 \psfrag{X}{Distance from the crack tip}
 \psfrag{Y}{\Large{$\sigma_{n}$}}
 \psfrag{b}{Fully developed bridging zone (FBZ) length}
 \psfrag{s}{$T_{n_{0}}$}
 \psfrag{p}{$A$}
 \psfrag{q}{$B$}
  \includegraphics[width=9in,height=6in]{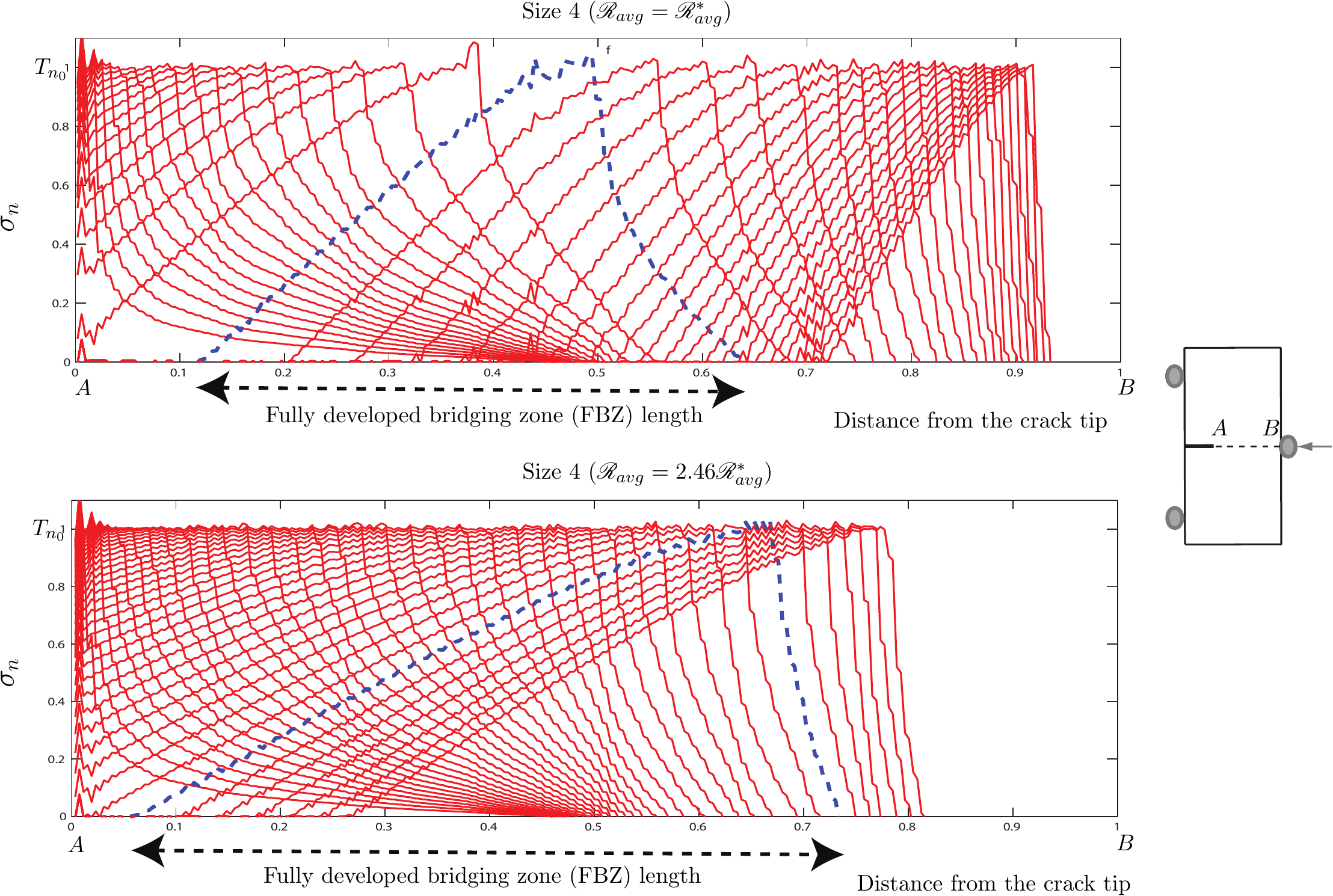}
  \caption[]{Snapshots of the normal stress $\sigma_{n}$ distribution along the crack path at different points in the loading cycle for Size-4 SETB specimen. The blue doted line (labeled `FBZ length') is the snapshot of the $\sigma_{n}$ distribution at the instance of formation of the FBZ which corresponds to the peak load in the loading cycle. `A' is the initial crack tip and only the positive ordinate axes is shown. These figures give an estimate of the bridging zone length, and help in understanding bridging zone formation, stabilization and movement.}
  \label{fig:bridgingSENB1}
\end{sidewaysfigure}

\end{document}